\documentclass[a4paper,twocolumn]{esapub} 
\usepackage{times}
\usepackage{natbib}
\usepackage{graphicx}
\usepackage{amsmath}
\usepackage{amssymb}

\def\apj{\rm{ApJ}}

\def\aap{\rm{A\&A}}

\def\mnras{\rm{MNRAS}}

\def\solphys{\rm{Solar~Phys.}}

\def\bu{\boldsymbol u}
\title{Stochastic excitation and damping of solar-type oscillations}
\author[]{G. Houdek}
\affil[]{Institute of Astronomy, University of Cambridge, 
          Cambridge CB3 0HA, UK}

\begin{document}


\maketitle

\begin{abstract}
A review on acoustic mode damping and excitation in solar-type stars is 
presented. Current models for linear damping rates are discussed in the light
of recent low-degree solar linewidth measurements with emphasis on the 
frequency-dependence of damping rates of low-order modes. Recent developments 
in stochastic excitation models are reviewed and tested against the latest
high-quality data of solar-like oscillations, such as from alpha Cen A, and
against results obtained from hydrodynamical simulations.
\end{abstract}

\vspace{-3mm}
\section{Introduction}
\vspace{-3mm}
Solar p modes are believed to be excited stochastically in the outer
layers of the star by the turbulent convection. Accurate measurements are
available from both space-borne (e.g., VIRGO, GOLF and MDI) and ground-based
(GONG and BiSON) observations which provide valuable data regarding the
excitation and damping processes influencing the modes.  
Such modes can be stochastically excited by the
turbulent convection. The process can be regarded as multipole acoustical 
radiation (e.g. Unno 1964). For solar-like stars, the acoustic
noise generated by convection in the star's resonant cavity 
may be manifest as an ensemble of p modes over a wide band in 
frequency (Goldreich \& Keeley 1977). The amplitudes are determined by 
the balance between the excitation and damping, and are expected to be 
rather low. The turbulent-excitation model predicts not 
only the right order of magnitude for the p-mode amplitudes (Gough 1980), 
but it also explains the observation that millions of modes are excited 
simultaneously. 
One particular observational detail, the frequency dependence of the
energy supply rate to acoustic modes, has been shown to be a
particularly important diagnostic property (e.g. Libbrecht 1988), 
for it has diagnostic potential for improving current stochastic-excitation
models (e.g., Balmforth 1992b, Chaplin et al. 2005).

The energy flow from radiation and convection
into and out of the p modes takes place very near the surface.
Sun-like stars possess surface convection zones, and it is in 
these zones, where the energy is transported principally by the 
turbulence, that most of the driving takes place.
Mode stability is governed not only by the perturbations in the radiative 
fluxes (via the $\kappa$-mechanism) but also by the perturbations in the 
turbulent fluxes (heat and momentum). The study of mode stability therefore 
demands a theory for convection that includes the interaction of the 
turbulent velocity field with the pulsation.

Our understanding of turbulent convection in stars is still only
rudimentary.  Simple phenomenological convection models based on 
mixing-length ideas are still the most widely used
for computing just the mean stratification of the convectively 
unstable layers. Attempts to describe the associated physics on 
all scales (both temporal and spatial) have led to similar simplified 
formalisms, which include one or more adjustable parameters that require 
calibration by high-quality observations.

\vspace{-3mm}
\section{Mode parameters}
\vspace{-3mm}
Were solar p modes to be genuinely linear and stable, their power spectrum
could be described in terms of an ensemble of intrinsically damped,
stochastically driven, simple-harmonic oscillators,
provided that the background
equilibrium state of the star were independent of time (Fig.\,1); 
if we assume further that mode phase fluctuations contribute negligibly to 
the width of the spectral lines, the intrinsic damping rates of
the modes, $\Gamma/2$, could then be determined observationally from 
measurements of the pulsation linewidths $\Gamma$.

The power (spectral density), $P$, of the surface displacement $\xi_{nl}(t)$ 
of a damped, stochastically driven, simple-harmonic oscillator, satisfying
\begin{equation}
I_{nl}\left[\frac{{\rm d}^2{\xi_{nl}}}{{\rm d}t^2}
+\Gamma_{nl}\frac{{\rm d}\xi_{nl}}{{\rm d}t}
+\omega_{nl}^2\xi_{nl}\right]=f(t)\,,
\label{eq:harmosc}
\end{equation}
which represents the pulsation mode of order $n$ and degree $l$, with 
linewidth $\Gamma_{nl}$ and frequency $\omega_{nl}$ and mode 
inertia $I_{\rm nl}$, satisfies 
\begin{equation}
P\propto P_{\rm L}\,P_{\rm f}=
\frac{\Gamma_{nl}/2\pi}
{(\omega-\omega_{nl})^2+\Gamma_{nl}^2/4}\,
P_{\rm f}\,,
\label{eq:powerdensity}
\end{equation}
assuming $\Gamma_{nl}\ll\omega_{nl}$, where 
$f(t)$ describes the stochastic forcing function.
Integrating equation~(\ref{eq:powerdensity}) over frequency leads to the
total mean energy in the mode
\begin{eqnarray}
I_{nl}\,V^2_{nl}\!\!\!\!&:=&\!\!\!\!
 \frac{1}{2}\omega^2_{nl}I_{nl}\langle|A_{nl}|^2\rangle
\propto\omega^2_{nl}I_{nl}\int_{-\infty}^\infty P(\omega)\,{\rm d}\omega\cr
&\propto&\frac{P_{\rm f}(\omega_{nl})}{\Gamma_{nl}}\,,
\label{eq:totalenergy}
\end{eqnarray}
where $A_{nl}$ is the displacement amplitude (angular brackets,
$\langle\rangle$, denote an expectation value) and $V_{n,l}$ is the 
rms velocity of the displacement. The total mean energy of a mode is therefore
directly proportional to the rate of work (also called excitation rate or 
energy supply rate) of the stochastic forcing $P_{\rm f}$ at the 
frequency $\omega_{nl}$ and indirectly proportional to $\Gamma_{nl}$.

\begin{figure}[t]
\centering
\includegraphics[width=0.9\linewidth]{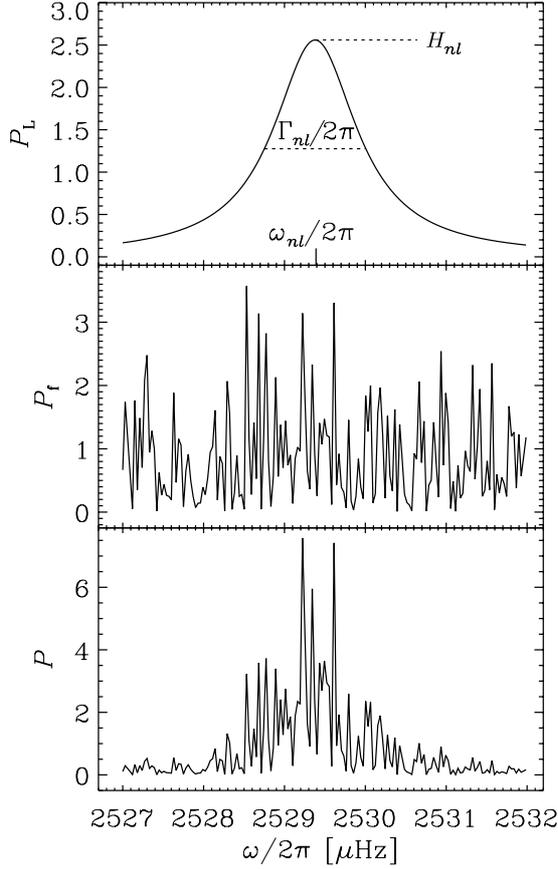}
\caption{
Power spectral density of a randomly excited, damped, harmonic oscillator. 
$P_{\rm f}$ represents the spectral density of the random force
and $P$ is the product of the Lorentzian $P_{\rm L}$ 
and $P_{\rm f}$ (adopted from Kosovichev 1995).
}
\end{figure}

Equations (\ref{eq:harmosc})--(\ref{eq:totalenergy}) are discussed in
terms of the displacement $\xi$ (from here on we omit the subscripts $n$ 
and $l$), but in order to have a direct relation between the observed
velocity signal $v(t)={\rm d}\xi/{\rm d}t$ and the modelled excitation 
rate $P_{\rm f}$ we shall first take the Fourier transform $\tilde V(\nu)$
of $v(t)$ ($\nu=\omega/2\pi$). It follows that the total mean energy $E$ 
of the harmonic signal of a single pulsation mode is then given by 
(Chaplin et al. 2005)
\begin{equation}
E=IV^2=I\hat\delta\int_{-\infty}^\infty\vert\tilde V(\nu)\vert^2\,{\rm d}\nu
=\frac{1}{4}I\Gamma H\,,
\label{eq:totalenergyH}
\end{equation}
in which
\begin{equation}
H:=\int_{\nu-\hat\delta/2}^{\nu+\hat\delta/2}\vert\tilde V(\nu)\vert^2\,{\rm d}\nu\,,
\label{eq:height}
\end{equation}
is the maximum power density - which corresponds to the `height' of the
resonant peak in the frequency domain (see Fig.~1). The height $H$ 
is the maximum of the discrete power, i.e. the integral of power spectral
density over a frequency bin $\hat\delta=1/T_{\rm obs}$, where $T_{\rm obs}$ 
is the total observing time. The following expressions
\begin{equation}
V^2\,:=\,\frac{P_{\rm f}}{\Gamma I}
\,=\,\frac{1}{4}\Gamma H
\label{eq:V-H}
\end{equation}
and
\begin{equation}
H\,:=\,\frac{P_{\rm f}}{(\Gamma/2)^2I}
\label{eq:H}
\end{equation}
provide a direct relation between the observed height $H$ 
(in cm$^2\,$s$^{-2}$Hz$^{-1}$), the modelled
energy supply rate $P_{\rm f}$ (in erg\,s$^{-1}$), and 
damping rate $\eta=\Gamma/2$.

\vspace{-3mm}
\section{Model computations}
\vspace{-3mm}
The model calculations need to solve the fully nonadiabatic pulsation
problem from which the damping rates are obtained from the imaginary
part $\eta=\omega_{\rm i}$ of the complex eigenfrequency
$\omega=\omega_{\rm r}+{\rm i}\omega_{\rm i}$.
Moreover, the pulsation dynamics depends crucially on the convection
treatment as reported by Balmforth (1992a) for the solar case and
by Houdek et al. (1999) for solar-type stars. Consequently the model 
computations should also take into account the effect of the turbulent 
fluxes (heat and momentum) on the pulsation properties (e.g., damping 
rates and shape of the eigenfunctions). 

Here we follow the calculations reported by Balmforth (1992a) and by 
Houdek et al. (1999).
In these calculations the equations describing the stellar structure 
and pulsations are
\begin{eqnarray}
\frac{\partial}{\partial m}(p_{\rm g}\!+\!p_{\rm t})
 \!+\!(3\!\!\!\!\!&-&\!\!\!\!\!\Phi)\frac{p_{\rm t}}{4\pi r^3\rho}\!=\!
 -\frac{1}{4\pi r^2}\left(\frac{Gm}{r^2}\!+\!\frac{\partial^2r}{\partial t^2}\right),\cr
\frac{\partial r}{\partial m}&=&\frac{1}{4\pi r^2\rho}\,,\cr
c_p\frac{\partial T}{\partial t}
  -\frac{\tilde\delta}{\rho}\frac{\partial p_{\rm g}}{\partial t}&=&
  -4\pi\frac{\partial}{\partial m}\left[r^2(F_{\rm r}\!\!+\!\!F_{\rm c})\right]\,,
\label{eq:structure}
\end{eqnarray}
resembling the conservation of momentum, mass and thermal energy. In 
equations~(\ref{eq:structure}) is $r$ the radius, $m$ is mass, $p_{\rm g}$ is 
gas pressure, $\rho$ is density, $T$ is temperature, $c_p$ is 
the specific heat at constant pressure, and 
$\tilde\delta:=-(\partial\ln\rho/\partial\ln T)_{p_{\rm g}}$.
The turbulent velocity field ${\bu}=(u_1,u_2,u_3)$ is described 
in a cartesian coordinate system and for a Boussinesq fluid, 
$p_{\rm t}=\langle\rho{u_3^2}\rangle$ is
the $(r,r)$-component of the Reynolds stress tensor (sometimes called the
turbulent pressure, and here the angular brackets denote an ensemble average). 
In the Boussinesq approximation the convective heat flux is 
$F_{\rm c}\simeq\rho c_p\langle{u_3\vartheta}\rangle$
($\vartheta$ being the convective temperature fluctuation), and
$\Phi=\langle{{\bu}\cdot{\bu}}\rangle/\langle{u_3^2}\rangle$
is a parameter that describes the anisotropy of the turbulent velocity field
${\bu}$. The radiative heat flux $F_{\rm r}$ is obtained from the
Eddington approximation to radiative transfer (Unno \& Spiegel 1966).\\
The equilibrium model is constructed from solving 
equations~(\ref{eq:structure}) but with the partial time derivatives set 
to zero, $\partial/\partial t=0$, which, for example, leads for the first 
integral of the thermal energy equation to the well-known expression 
$L(r)=4\pi r^2(F_{\rm r}+F_{\rm c})$.

The pulsation equations for radial modes are then obtained from linear
perturbation theory
\begin{eqnarray}
\frac{\partial}{\partial m}\left(\frac{\delta p}{p}\right)&=&
 \hat f\left(\frac{\delta r}{r},\frac{\delta T}{T},\frac{\delta p}{p},\frac{\delta p_{\rm t}}{p},\frac{\delta\Phi}{\Phi}\right)\,,\cr
\frac{\partial}{\partial m}\left(\frac{\delta r}{r}\right)&=&
-\frac{1}{4\pi r^3\rho}\left(3\frac{\delta r}{r}+\frac{\delta\rho}{\rho}\right)\,,\cr
\frac{\partial}{\partial m}\left(\frac{\delta L}{L}\right)&=&
-{\rm i}\omega\frac{c_pT}{L}\left(\frac{\delta T}{T}-\nabla_{\rm ad}\frac{\delta p}{p}\right)\,,
\label{eq:pulsation}
\end{eqnarray}
where $\delta$ is the Lagrangian perturbation operator, and for simplicity the
right hand side of the perturbed momentum equation is formally expressed by the
function $\hat f$ (the full set of equations can be found in, e.g., 
Balmforth 1992a).
Equations~(\ref{eq:pulsation}) are solved subject to boundary conditions to 
obtain the eigenfunctions and the complex angular eigenfrequency 
$\omega=\omega_{\rm r}+{\rm i}\eta$, where $\omega_{\rm r}$ is the (real) 
pulsation frequency and $\eta=\Gamma/2$ is the damping rate in (s$^{-1}$).  
The turbulent flux perturbations of heat and momentum, $\delta L_{\rm c}$ and
$\delta p_{\rm t}$, and the fluctuating anisotropy factor $\delta\Phi$ are 
obtained from the nonlocal, time-dependent convection formulation 
by Gough (1977a,b). 

\vspace{-3mm}
\section{Amplitude ratios and phases in the solar atmosphere}
\vspace{-3mm}
A useful test of the pulsation theory, independent of an excitation model,
can be performed by comparing the theoretical intensity-velocity amplitude
ratios
\begin{equation}
\frac{\Delta L_{\rm s}}{\Delta V}:=
                        \frac{\delta L/L}{\omega_{\rm r} r\;\delta r/r}
\label{eq:amprat}
\end{equation}
with observations.  
It is, however, important to realize that various instruments observe in
different absorption lines and consequently at different heights in the 
atmosphere. This property has to be taken into account not only when comparing
observations between various instruments (e.g. Christensen-Dalsgaard \& Gough 1982, 
see also Bedding, these proceedings), but also when comparing theoretical 
amplitude estimates with observations (Houdek et al. 1995). Table~1 
lists some of the relevant properties of various instruments.
\begin{table}
\begin{center}
\caption{Absorption lines and their wavelengths $\lambda$ of various 
helioseismic instruments. Also listed are the optical depths $\tau_{5000}$ 
at 5000\,\AA\ and the corresponding approximate heights above the photosphere 
($h=0$ at $T=T_{\rm eff}$) at which the lines are formed.
}
\bigskip
\renewcommand{\arraystretch}{1.2}
\begin{tabular}[t]{lcclc}
Instrument$\!\!\!$&$\!\!\!\!\!\!\!$line$\!\!\!\!\!\!$&$\lambda\,$(\AA)&$\tau_{5000}$&$\!\!\!\!\!\!$height\,(km)$\!\!\!\!$\\
\noalign{\medskip}
\hline
\noalign{\medskip}
BBSO &Ca      &6439&0.05            &$\sim129^a$\\
BiSON&K       &7699&0.013           &$\sim250^b$\\
MDI  &Ni\,I   &6708&9$\times10^{-3}$&$\sim300^c$\\
GOLF &Na D1/D2&5690&5$\times10^{-4}$&$\sim500^c$\\
\end{tabular}
\end{center}
\hbox{\hspace{0pt} $^a$from Libbrecht (1988),\hfil} 
\hbox{\hspace{0pt} $^b$from Christen-Dalsgaard \& Gough (1982),\hfil}
\hbox{\hspace{0pt} $^c$from Toutain et al. (1997, but see also\hfil}
\hbox{\hspace{0pt} \phantom{$^c$from\ }Baudin et al., these proceedings).}
\vspace{-2mm}
\end{table} 

\begin{figure}[t]
\centering
\includegraphics[width=0.96\linewidth]{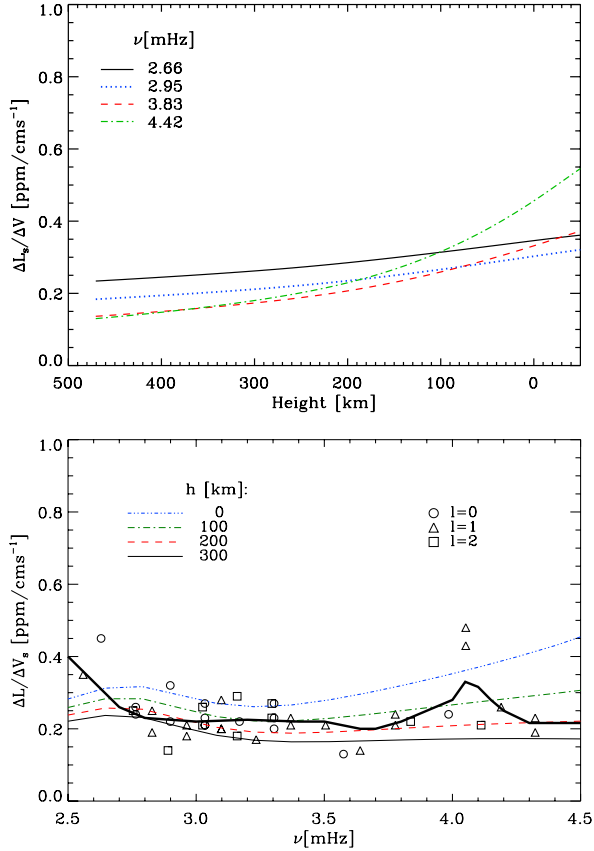}
\caption{
Top:  
Calculated amplitude ratios (see equation (\ref{eq:amprat})) as a function 
of height in a solar model for modes with different frequency values.
Bottom: 
Theoretical amplitude ratios (surface luminosity perturbation over velocity)
for a solar model compared with observations by Schrijver et\,al. (1991). 
Computed results are depicted at different heights above the photosphere
($h$=0\,km at $T=T_{\rm eff}$) The thick, solid curve indicates a 
running-mean average of the data (from Houdek et~al. 1995).
} 
\label{fig:amprat}
\end{figure}

In the top panel of Fig.~\ref{fig:amprat} the theoretical amplitude ratios
(equation~(\ref{eq:amprat})) of a solar model are plotted as function of 
height for several radial pulsation modes. The mode energy density
(which is proportional to $r\rho^{1/2}\delta r$) increases rather slowly with 
height; the density $\rho$, however, decreases very rapidly and consequently 
the displacement eigenfunction $\delta r$ increases with height. This leads 
to the results shown in the upper panel of Fig.~\ref{fig:amprat} where the
decrease in the amplitude ratios with height is particularly pronounced for
high-order modes for which the eigenfunctions vary rapidly in the evanescent 
outer layers of the atmosphere. It is for that reason why solar velocity 
amplitudes from, e.g., the GOLF instrument have larger values than the
measurements from the BiSON instrument (by about 25\%, Kjeldsen et~al 2005).\\
The lower panel of Fig.~\ref{fig:amprat} compares the estimated solar amplitude 
ratios (curves) with observed ratios (symbols) as function of frequency. The 
model results are depicted for velocity amplitudes computed at different 
atmospheric levels. The observations are obtained from accurate irradiance
measurements from the IPHIR instrument of the PHOBOS 2 spacecraft with
contemporaneous low-degree velocity data from the BiSON instrument at
Tenerife (Schrijver et al. 1991). The thick solid curve represents a 
running-mean average, with a width of 300$\,\mu$Hz, of the observational data.
The theoretical ratios for $h=200\,$km (dashed curve) show reasonable 
agreement with the observations.

\begin{figure}[t]
\centering
\includegraphics[width=1.00\linewidth]{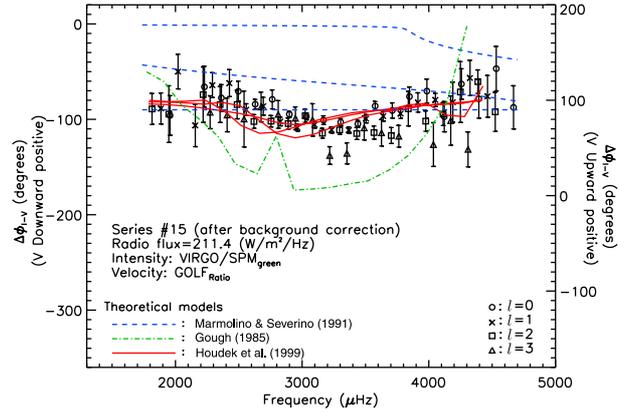}
\caption{ 
Intensity-velocity phase differences of low-degree p modes at the maximum of 
solar activity. Symbols are the results from intensity observations in the 
green channel of the VIRGO/SPM instrument on the SOHO spacecraft, and from
contemporaneous velocity observations from the GOLF instrument. Theoretical
results are depicted for three models: the dashed curves assume Newton's 
cooling law for energy transfer with three different values for the 
damping times,
the dot-dashed curve includes a local treatment of convection dynamics and
the diffusion approximation to radiative transfer, and the 
solid curves are the results from the model discussed
in Section~3 for three different sets of convection parameters
(adopted from Jim\'enez 2002).
}
\label{fig:phases}
\end{figure} 

An additional useful property of the oscillations is the phase difference
between intensity and velocity. Phase shifts between intensity observations
from the VIRGO/SPM instrument and contemporaneous velocity measurements from 
the GOLF instrument (symbols with error bars) are plotted as function of 
frequency in Fig.~\ref{fig:phases} together with results 
from various model computations (curves). The dashed curves are the 
results from calculations by Marmolino \& Severino (1991) who treated
the energy transport with Newton's cooling law assuming various values
for the damping time; the dot-dashed curve
is the result by Gough (1985) who used the diffusion approximation to
radiative transfer and a local, time-dependent mixing-length model to estimate
the turbulent fluxes. The solid curves are the results from the model
discussed in Section~3 and for three different sets of convection 
parameters (Houdek\;et\;al.\;1999). The latter results are in good 
agreement with the observations.\\
The results depicted in Figs~\ref{fig:amprat} and \ref{fig:phases} provide 
us with confidence that the
pulsation computations reproduce the properties of the radial eigenfunctions
in the outer atmospheric solar layers reasonably well.
\vspace{-4mm}
\section{Solar damping rates}
\vspace{-4mm}
Damping of stellar oscillations arises basically from two sources: processes 
influencing the momentum balance, and processes influencing the thermal energy 
equation. Each of these contributions can be divided further according to their
physical origin, which was discussed in detail by Houdek et al. (1999).

Important processes that influence the thermal energy balance are
nonadiabatic processes attributed to the modulation of the convective heat
flux by the pulsation. This contribution is related to the way that convection
modulates large-scale temperature perturbations induced by the pulsations
which, together with the conventional $\kappa$-mechanism, influences 
pulsational stability.\\
Current models suggest that an important contribution that influences
the momentum balance is the exchange of energy between the pulsation and 
the turbulent velocity field through dynamical effects of the fluctuating 
Reynolds stress. In fact, it is the modulation of the turbulent fluxes by 
the pulsations that seems to be the predominant mechanism responsible for 
the driving and damping of solar-type acoustic modes.
It was first reported by Gough (1980) that the dynamical effects arising 
from the turbulent momentum flux perturbation $\delta p_{\rm t}$ 
contribute significantly to the damping $\Gamma$. 
Detailed analyses (Balmforth 1992a) reveal how damping is controlled largely 
by the phase difference between the momentum perturbation and the density 
perturbation. Therefore, turbulent pressure fluctuations must not be neglected 
in stability analyses of solar-type p modes. 
\begin{figure*}[t]
\centering
\includegraphics[width=1.00\linewidth]{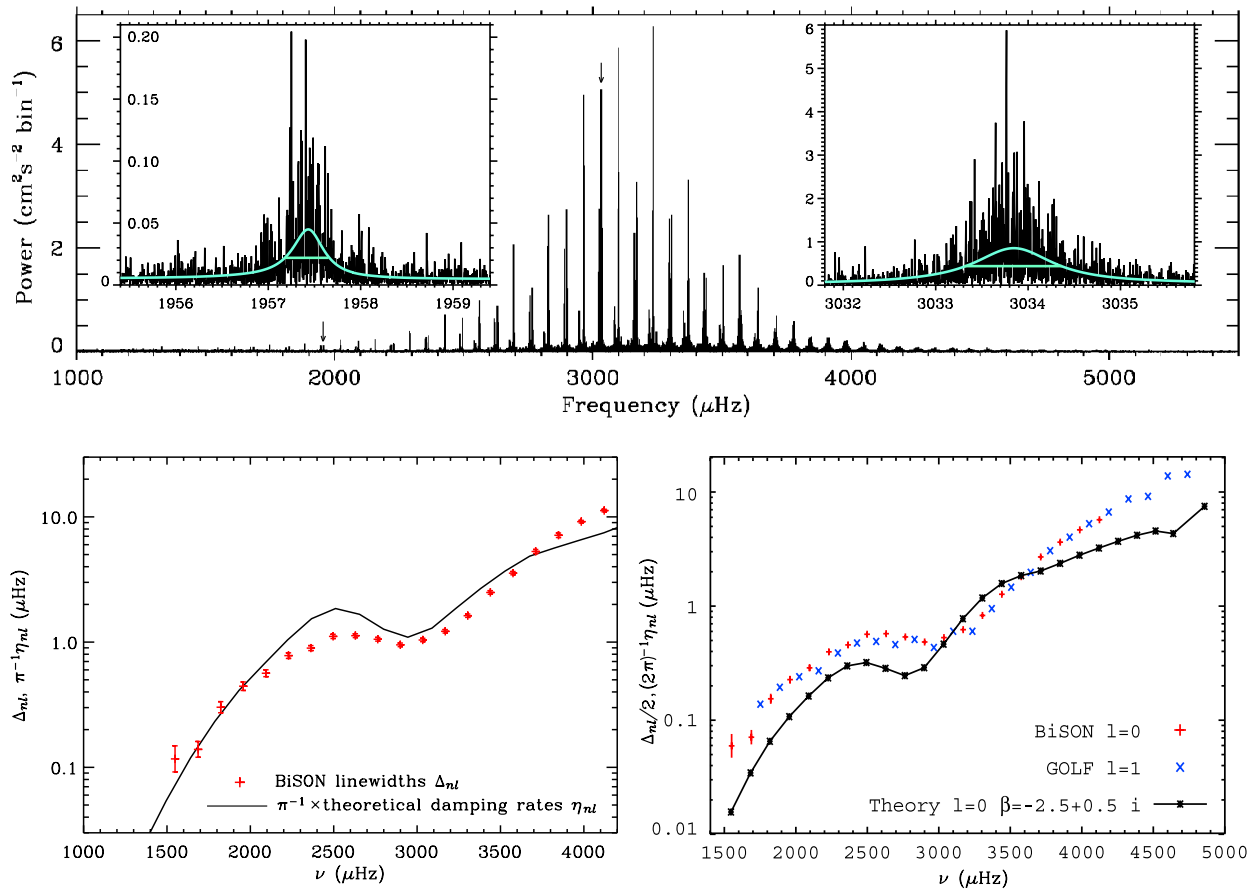}
\caption{
Top: Power spectrum of solar low-degree p modes obtained from a 3456-d
data set collected by BiSON between 1991 and 2000 (Chaplin et al. 2005). 
The two insets show Lorentzian profile fits (and their full-width at 
half-maximum (FWHM); solid, blue curves) to the spectral peaks of 
radial modes with order $n=13$ (left) and $n=21$ (right);
both spectral peaks are indicated by vertical arrows in the power spectrum. 
Bottom: The symbols in the left-hand panel are the measured BiSON linewidths
$\Delta_{nl}=\Gamma_{nl}/2\pi$ (we denote the FWHM in unit of cyclic 
frequency by $\Delta_{nl}$) which are 
compared with the theoretical damping rates $\pi^{-1}\eta_{nl}$ (connected 
by the solid curve) obtained from the model computations of Section~3 (from
Chaplin et al. 2005). In the right-hand panel theoretical results of
$(2\pi)^{-1}\eta_{nl}$ (solid curve) by Dupret et al. (2004) are compared 
with observations of $\Delta_{nl}/2$ (symbols).
} 
\label{fig:damping}
\end{figure*} 

A comparison between the latest linewidth measurements (full-width at 
half-maximum) $\Delta_{nl}=\Gamma_{nl}/2\pi$ and theoretical damping 
rates is given in Fig.~\ref{fig:damping}. The observational time series
from BiSON (Chaplin et al. 2005) was obtained from a 3456-d data set and
the linewidths of the temporal power spectrum (top panel) extend over many
frequency bins $\hat\delta=1/2T_{\rm obs}$. In that case the linewidth in 
units of cyclic frequency is related to the damping rate according to
\begin{equation} 
\Delta_{nl}=\pi^{-1}\eta_{nl}\,.
\end{equation} 
Recently Dupret et al. (2004, see also these proceedings) performed similar 
stability computations for the Sun using the time-dependent mixing-length 
formulation by Gabriel et al. (1975, 1998) which is based on the formulation 
by Unno (1967). The outcome of their computations is illustrated in the lower
right panel of Fig.~\ref{fig:damping}, which also shows the characteristic
plateau near 2.8\,mHz. It is, however, interesting to note that their findings
suggest the fluctuating convective heat flux to be the main contribution 
to mode damping, whereas for the model results shown in the lower left panel of 
Fig.~\ref{fig:damping}, which are based on Gough's (1977a,b) convection 
formulation, it is predominantly the fluctuating Reynolds stress that 
makes all modes stable.  
\vspace{-3mm}
\section{Excitation model and solar amplitudes}
\vspace{-3mm}
Because of the lack of a complete model for convection, the mixing-length
formalism still represents the main method for computing the turbulent
fluxes in the convectively unstable layers in a star. One of the assumptions
in the mixing-length formulation is the Boussinesq approximation, which
results in neglecting the acoustic wave generation by assuming the fluid to
be incompressible. Consequently a separate model is needed to estimate the
rate of the acoustic noise (energy supply rate) generated by the turbulence.
The excitation process can be regarded as multipole acoustic radiation
(Lighthill 1952).
Acoustic radiation by turbulent multipole sources in the context of stellar
aerodynamics has been  considered by Unno \& Kato (1962), Moore \& 
Spiegel (1964), Unno (1964), Stein (1967),  Goldreich \& Keeley (1977), 
Bohn (1984), Osaki (1990), Goldreich \& Kumar (1990), Balmforth (1992b), 
Goldreich, Murray \& Kumar (1994), Musielak et~al. (1994), 
Samadi \& Goupil (2001) and Chaplin et~al. (2005).
\vspace{-4mm}
\subsection{Excitation model}
\vspace{-3mm}
The mean amplitude $A$ of a mode is determined by a balance by the
energy supply rated $P_{\rm f}$ from the turbulent velocity field
and the thermal and mechanical dissipation rate characterized by
the damping coefficient $\eta$.
The procedure that we adopt to estimate $A$ is that of Chaplin et~al. (2005),
whose prescription follows that of Balmforth (1992b).

As in Section~2 we represent the linearized pulsation dynamics by the 
simplified equation
\begin{equation}
\rho\left(\frac{\partial^2\boldsymbol\xi}{\partial t^2}
         +2\eta\frac{\partial\boldsymbol\xi}{\partial t}
         +\mathcal{L}\boldsymbol\xi\right)=
\boldsymbol{\mathcal{F}}(\bu)+\boldsymbol{\mathcal{G}}(s^{\prime})
\label{eq:excitation}
\end{equation}
for the displacement $\boldsymbol\xi(\boldsymbol r,t)$, which is now
also a function of radius $\boldsymbol r$, of a forced oscillation 
corresponding to a single radial mode satisfying the homogeneous equation
\begin{equation}
{\cal L}\boldsymbol\xi(\boldsymbol r)=
\omega^2\boldsymbol\xi(\boldsymbol r)
\end{equation}
in which $\omega$ (and $\boldsymbol\xi$) are real and ${\cal L}$ is
a linear spatial operator. The (inhomogeneous) fluctuating terms on
the right-hand-side of equation~(\ref{eq:excitation}) arise from the
fluctuating Reynolds stresses
\begin{equation}
\boldsymbol{\cal F}(\bu)=\nabla\cdot(\rho\bu\bu-
\langle\rho\bu\bu\rangle)
\label{eq:F}
\end{equation}
and from the fluctuating gas pressure (due to the fluctuating buoyancy force),
represented by $\boldsymbol{\mathcal{G}}(s^{\prime})$, where $s^{\prime}$ 
is the Eulerian entropy fluctuation (Bohn 1984; Osaki 1990; 
Goldreich \& Kumar 1990; Balmforth 1992b; 
Goldreich, Murray \& Kumar 1994; Samadi \& Goupil 2001).
The latest numerical simulations by 
Stein et al. (2004) suggest that both forcing terms in 
equation~(\ref{eq:excitation}) contribute to the energy supply rate $P_{\rm f}$ 
by about the same amount, a result that 
was also reported by Samadi et al. (2003) using the turbulent velocity 
field and anisotropy factors from numerical simulations
(Stein \& Nordlund 2001).
In this paper we consider only the term of the fluctuating Reynolds stresses
and because we use only radial modes, only the vertical
component of $\boldsymbol{\cal F}$ is important,
\begin{equation}
{\cal F}_3(u_3)\simeq
\frac{\partial}{\partial r}(\rho u^2_3-\langle\rho u^2_3\rangle)\,.
\label{eq:F3}
\end{equation}
If we define the vertical component of the velocity correlation as
$R_{33}=\langle u_3u_3\rangle$, its Fourier transform $\widehat R_{33}$ can
be expressed in the Boussinesq-quasi-normal approximation 
(e.g. Batchelor 1953) as a function of the turbulent energy spectrum function
$E(k,\omega)$: 
\begin{equation}
\widehat R_{33}=\frac{\Psi E(k,\omega)}{12\pi k^2}\,,
\end{equation}
where $k$ is a wavenumber and $\Psi$ is an anisotropy parameters given by
\begin{equation}
\Psi=\left[{2\Phi}/{3(\Phi-1)}\right]^{1/2}
\end{equation}
which is unity for isotropic turbulence (Chaplin et~al. 2005). This factor
was neglected in previously published excitation models but it has to be 
included in a consistent computation of the acoustic energy supply rate. 
Following Stein (1967) we factorize the energy spectrum function into
$E(k,\omega)=E(k)\Omega(\omega; \tau_k)$, where $\tau_k=\lambda/ku_k$ 
is the correlation time-scale of eddies of size $\pi/k$ and velocity $u_k$;
the correlation factor $\lambda$ is of order unity and accounts for
uncertainties in defining $\tau_k$.
\begin{figure}[t]
\centering
\includegraphics[width=0.95\linewidth]{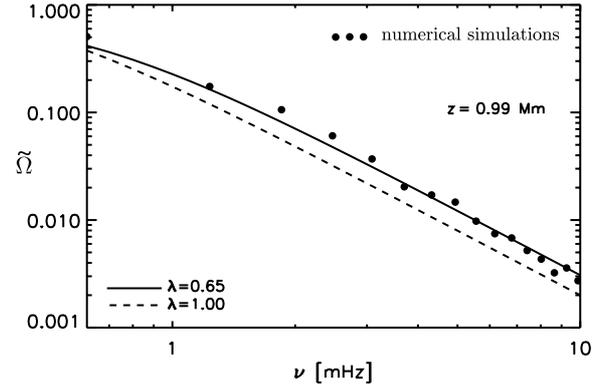}
\caption{ 
Comparison of the frequency factor $\tilde\Omega(\omega;\tau_k)$ 
(see equation~(\ref{eq:function-S})) between hydrodynamical 
simulation results for solar convection
(Stein \& Nordlund 2001; symbols) and the Lorentzian frequency factor 
by Samadi et al. (2003) for two different values of the correlation 
factor $\lambda$ (solid and dashed curves). The results are plotted
as function of frequency evaluated 0.99\,Mm below the solar photosphere 
(adopted from Samadi et~al. 2003). 
} 
\label{fig:Lorentzian}
\end{figure} 

\begin{figure}[t]
\centering
\includegraphics[width=1.00\linewidth]{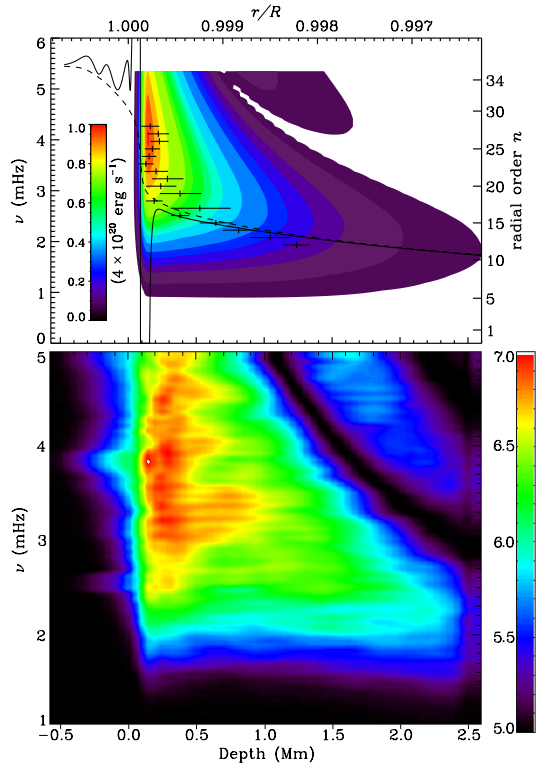}
\caption{
Top: 
The contours represent the integrand of the acoustic energy supply rate 
(\ref{eq:excitation-rate}), multiplied by 
$I^{-1}{\rm d}r/{\rm d}\ln p$, and are plotted
as function of frequency $\nu$ and the depth variable 
$z={\rm R}_\odot\!\!-\!r$.
The plus symbols with (horizontal) error bars
are the measured locations of the excitation regions by
Chaplin \& Appourchaux (1999). 
The solid and dashed curves denote the exact acoustic cutoff
frequency and the approximation in an isothermal region, $c/2H_p$,
where $c$ is the sound speed and $H_p$ is the pressure scale height.
Bottom: 
Logarithm of integrand of the acoustic energy supply rate from
Stein \& Nordlund's (2001) hydrodynamical simulations. 
} 
\label{fig:excitation-integrand}
\end{figure} 
The energy supply rate is then given by (see Chaplin et~al. 2005 for details)
\begin{equation}
P_{\rm f}=
    \frac{\pi}{9I}
    \int_0^R \ell^3
    \left(\Phi\Psi rp_{\rm t}\frac{\partial{\xi}_{r}}{\partial r}\right)^2
    {\cal S}(r;\omega)\,{\rm d}r\,,
\label{eq:excitation-rate}
\end{equation}
with
\begin{equation}
{\cal S}(r;\omega)=\int_0^\infty \kappa^{-2}\tilde E^2(\kappa)
                     \tilde\Omega(\tau_k;\omega)\,{\rm d}\kappa\,,
\label{eq:function-S}
\end{equation}
where $\kappa=k\ell/\pi$, $\ell$ is the mixing length, $R$ is surface radius,
and $\xi_r$ is the normalized radial part of $\boldsymbol{\xi}$.
The spectral function ${\cal S}$ accounts for contributions to $P_{\rm f}$
from the small-scale turbulence and includes the normalized spatial 
turbulent energy spectrum $\tilde E(k)$ and the 
frequency-dependent factor $\tilde\Omega(\tau_k;\omega)$. For $\tilde E(k)$
it has been common to adopt either the Kolmogorov (Kolmogorov 1941)
or the Spiegel spectrum (Spiegel 1962). The frequency-dependent factor
$\tilde\Omega(\tau_k;\omega)$ is still modelled in a very rudimentary way
and we adopt two forms:\\
-- the Gaussian factor (Stein 1967)\,,
\begin{equation}
\tilde\Omega_{\rm G}(\omega;\tau_k)=
    \frac{\tau_k}{\sqrt{2\pi}}\,{\rm e}^{-(\omega\tau_k/\sqrt2)^2}\,;
\label{eq:Gaussian}
\end{equation}
-- the Lorentzian factor 
   (Gough 1977b; Samadi et al. 2003; Chaplin et al. 2005)\,,
\begin{equation}
\tilde\Omega_{\rm L}(\omega;\tau_k)=
 \frac{\tau_k}{\pi\sqrt{2\ln2}}\,\frac{1}{1+(\omega\tau_k/\sqrt{2\ln2})^2}\,.
\label{eq:Lorentzian}
\end{equation}
The Lorentzian frequency factor is a result predicted for the largest, 
most-energetic eddies by the time-dependent mixing-length formulation 
of Gough~(1977b).  Recently, Samadi et al. (2003) reported that 
Stein \& Nordlund's hydrodynamical simulations also suggest a Lorentzian 
frequency factor (see Fig.~\ref{fig:Lorentzian}), which decays more slowly
with depth $z$ and frequency $\omega$ than the Gaussian factor. Consequently
a substantial fraction to the integrand of 
equation~(\ref{eq:excitation-rate}) arises from eddies situated in the deeper
layers of the Sun, resulting in a larger acoustic excitation 
rate $P_{\rm f}$.
\begin{figure}[t]
\centering
\includegraphics[width=0.95\linewidth]{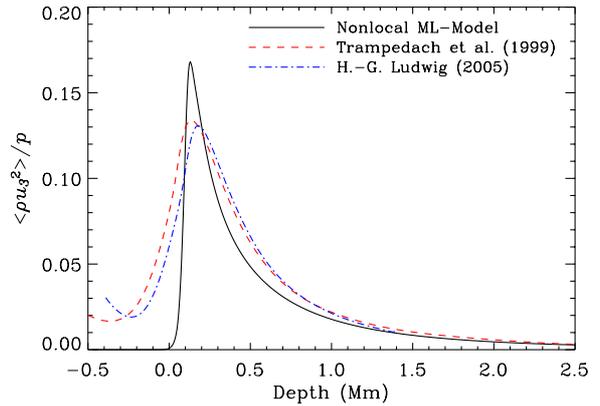}
\caption{Reynolds stress as a function of the depth variable 
$z={\rm R}_\odot-r$ for 
various solar models. Results are shown for the nonlocal mixing-length model
(solid, black curve) and from hydrodynamical simulations by 
Trampedach et~al. (1999, dashed, red curve) and Ludwig (2005, dot-dashed, blue
curve). 
} 
\label{fig:reynolds-stress}
\end{figure}

Another way to study the properties of the turbulent-excitation model is to
compare the depth and frequency dependence of the integrand of $P_{\rm f}$ 
with solar measurements and with hydrodynamical simulations of the Sun.
The results are depicted in Fig.\,\ref{fig:excitation-integrand}: the 
contours in the top panel show the integrand of 
equation~(\ref{eq:excitation-rate}) multiplied by 
$I^{-1}$d$r$/d$\ln p$.
The contours are compared with measurements of the locations of the driving
regions by Chaplin \& Appourchaux's (1999), represented by the plus symbols
with (horizontal) error bars.
The modelled locations of the excitation regions are in good agreement with 
the observations, particularly for modes with frequencies larger than the 
acoustic cutoff frequency (these modes are therefore propagating). 
The cutoff frequency is the solid curve and the dashed curve is the isothermal 
approximation $c/2H_p$, where $c$ is the sound speed and $H_p$ is the 
pressure scale height. For these (propagating) modes the frequency dependence 
of $P_{\rm f}$ is predominantly determined by the turbulent spectrum 
(represented by the spectral function 
${\cal S}$, cf. equation~(\ref{eq:function-S})). Modes with 
frequencies smaller than the acoustic cutoff frequency are evanescent, 
and their frequency dependence of $P_{\rm f}$ is predominantly 
determined by the shape of the eigenfunctions, i.e. by the term 
$(\partial\xi_r/\partial r)^2/I$, and consequently by the structure of the 
equilibrium model.\\
\begin{figure*}[t]
\centering
\includegraphics[width=1.00\linewidth]{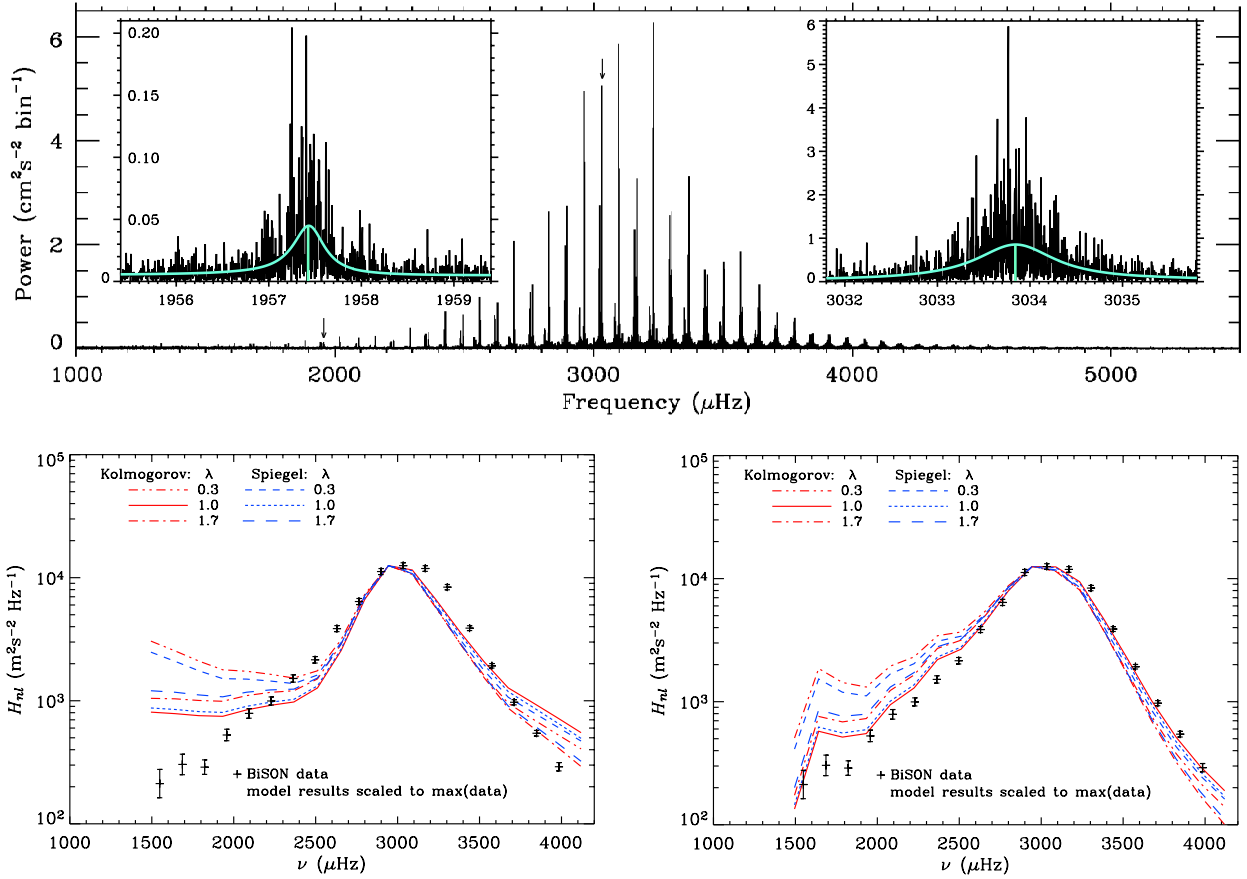}
\caption{
Top: 
Power spectrum of solar low-degree p modes obtained from a 3456-d
data set by BiSON (Chaplin et al. 2005). 
The two insets show Lorentzian profile fits (and their heights $H$; 
solid, blue curves) to the spectral peaks of radial modes with order 
$n=13$ (left) and $n=21$ (right); both spectral peaks are indicated
by vertical arrows in the power spectrum. 
Bottom:
Radial mode peak heights, $H$, from
observation and modelling.  The crosses (with error bars) are
well determined heights extracted from fits to the 3456-d BiSON
spectrum. The curves show predicted heights $H$ for two
turbulent spectra, due to Kolmogorov and Spiegel, with a Gaussian factor
$\Omega_{\rm G}$, and assuming various values for the eddy correlation factor 
$\lambda$; they are scaled such
that their maximum values agree with the maximum value of the data.
Left-hand panel -- the theoretical curves were computed using
modelled values for both the forcing $P_{\rm f}$ and linear damping 
rates $\eta$. 
Right-hand panel -- the theoretical curves were computed using
the modelled values for the forcing $P_{\rm f}$ but with the damping 
rates replaced by $\pi\Delta$, where $\Delta$ are observed 
mode widths.
} 
\label{fig:heights}
\end{figure*} 
The lower panel of Fig.\,\ref{fig:excitation-integrand} shows the logarithm 
to the base 10 of the integrand of the excitation rate of the
numerical simulations by Stein \& Nordlund~(2001). As for the model
results (top panel), the extent of the driving regions decreases
with frequency; however, the simulations show broader excitation regions than 
the model results. This is predominantly a result of the different spatial
extents of the Reynolds stresses $\langle\rho u^2_3\rangle$ between the 
hydrodynamical simulations and the nonlocal mixing-length model. A plot
of $\langle\rho u^2_3\rangle/p$ ($p$ is the total pressure) as a function of
the depth variable $z={\rm R}_\odot-r$ is illustrated in 
Fig.~\ref{fig:reynolds-stress}. The Reynolds stress of the nonlocal 
mixing-length model shows a narrow peak near $z\simeq120\,$km and falls off
more rapidly with $z$ than the results from both hydrodynamical simulations.
This leads to a smaller excitation rate for the mixing-length model compared
with the hydrodynamical results and consequently the modelled heights $H$ need
to be scaled with a scaling factor $\Lambda>1$ in order to reproduce the 
observed values of the mode peak heights (Chaplin et al. 2005).
\begin{figure}[t]
\centering
\includegraphics[width=1.00\linewidth]{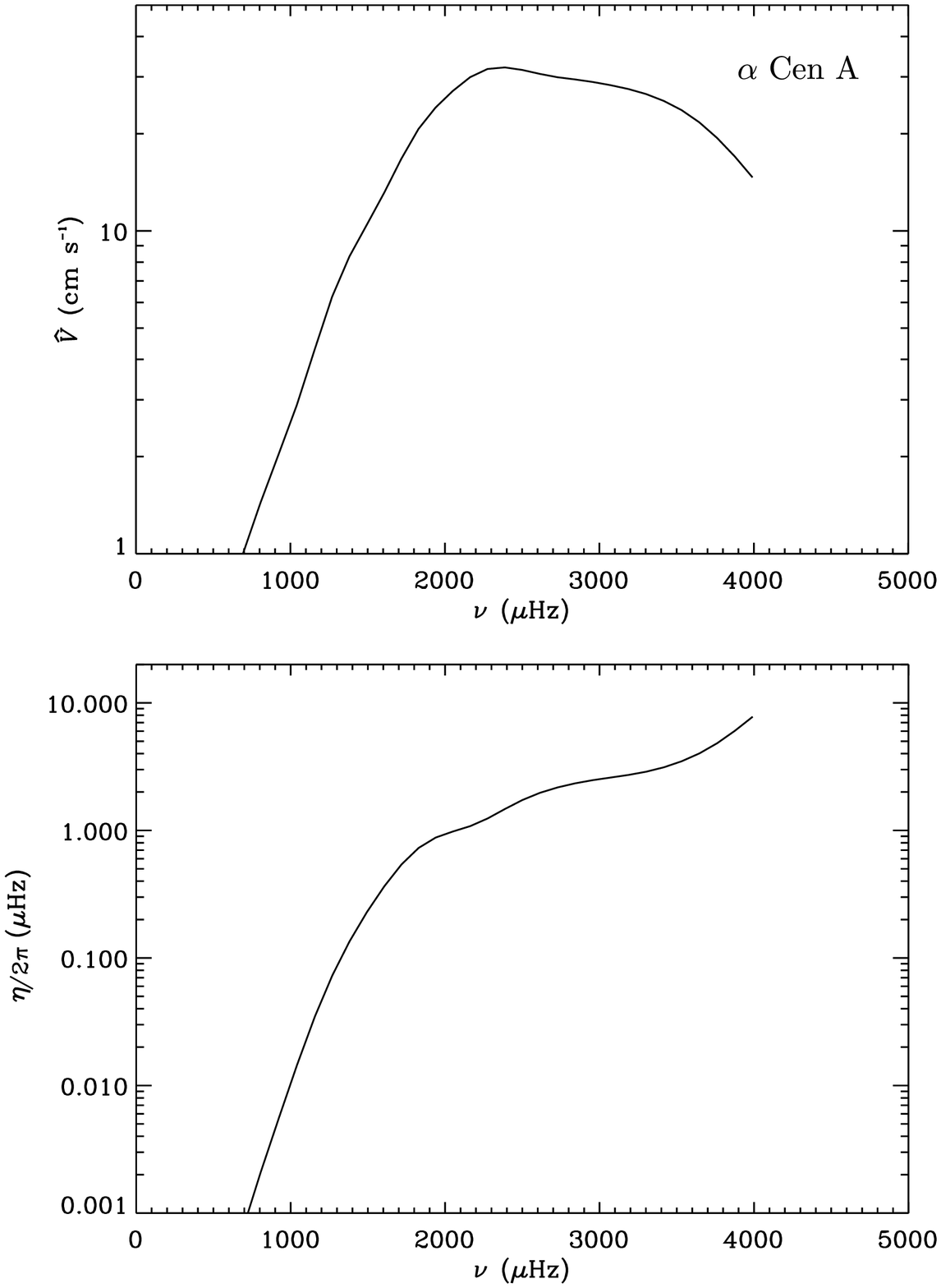}
\caption{
Top:
Predicted apparent velocity amplitudes (defined to be $\sqrt2$ times the 
rms value) for a model of $\alpha$~Cen~A, computed according to 
equation~(\ref{eq:V-H}).  
Bottom:
Linear damping rates for a model of $\alpha$~Cen~A, obtained by solving 
the fully nonadiabatic pulsation equations~(\ref{eq:pulsation}).
} 
\label{fig:aCenA}
\end{figure} 
\begin{figure}[t]
\centering
\includegraphics[width=1.00\linewidth]{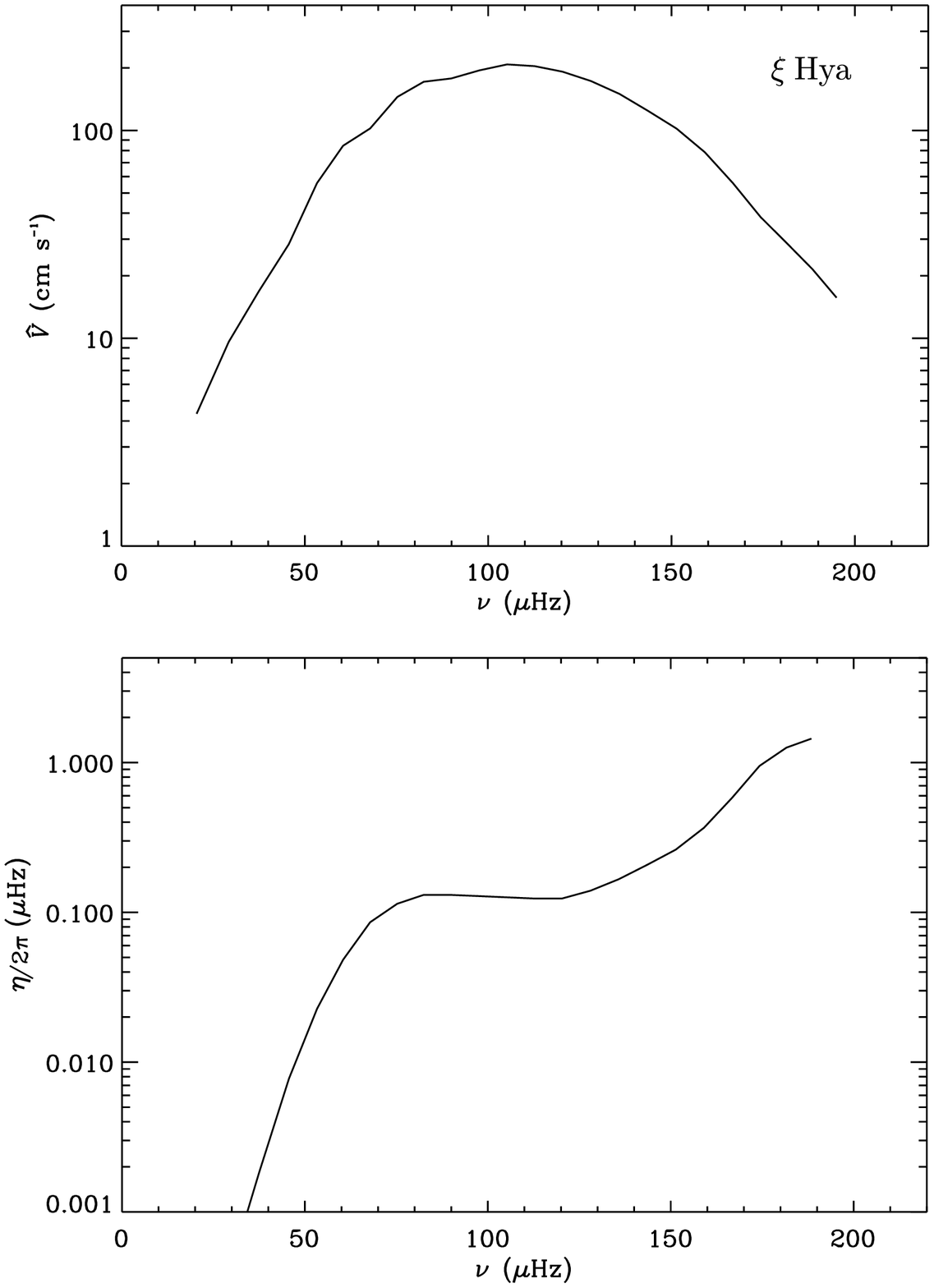}
\caption{
Top: 
Predicted apparent velocity amplitudes (defined to be $\sqrt2$ times the 
rms value) for a model of $\xi$~Hydrae, computed according to 
equation~(\ref{eq:V-H}).
Bottom:
Linear damping rates for a model of $\xi$~Hydrae, obtained by solving the 
fully nonadiabatic pulsation equations~(\ref{eq:pulsation}) 
(adopted from Houdek \& Gough 2002).
}
\label{fig:xiHya}
\end{figure} 
\vspace{-4mm}
\subsection{Solar amplitudes}
\vspace{-3mm}
Model predictions of the mode height $H$ were computed according to 
equation~(\ref{eq:H}) assuming $\Phi$ to be constant; the value 1.13
was adopted because it produces the best fit to the linewidth data
(cf. lower left panel of Fig.~\ref{fig:damping}). The resulting values are
joined by lines of various style in the lower panels of 
Fig.~\ref{fig:heights} for different 
values of the correlation factor $\lambda$ and for the two different
forms of the spatial energy spectrum $\tilde E(k)$, in all cases with the
Gaussian frequency factor given by equation~(\ref{eq:Gaussian}).
The BiSON measurements are rendered as crosses (with error bars). 
The maximum values of the theoretical results have been scaled
to the maximum observed value.
The values of the scaling factor $\Lambda$ by which we have multiplied
the theoretical values are given in Chaplin et~al. (2005).
The scaled theoretical estimations of $H$ shown in the lower left-hand panel
of the figure used modelled values of both the energy supply rate $P_{\rm f}$
and the linear damping rate $\eta$. They exceed the observed values 
at both low and high frequency, largely as a consequence of 
the modelled linewidths being larger than their measured counterparts
at intermediate frequencies. This is demonstrated by comparison with the
lower right-hand panel of Fig.~\ref{fig:heights}, where predictions of $H$
have been made using the modelled energy supply rate but with the
observed BiSON linewidths $\Delta$, rather than the modelled
damping rates $\eta$. There is a dramatic improvement in the modelled 
spectral heights $H$,
although agreement with observations remains imperfect. The discrepancy 
that remains when using observed linewidths is indicative of error in 
the equilibrium model and in the modelling of the energy supply rate 
$P_{\rm f}$.
This comparison reveals that the principal error in the theory, 
at least for the Sun, rests in the calculation of the damping rates,
particularly for low-order modes. These modes have upper turning points 
well below the region where most of the excitation takes place.\\
Chaplin et~al. (2005) have also computed mode heights $H$ using the Lorentzian 
frequency factor~(\ref{eq:Lorentzian}), but have found that in that case the 
heights at low frequency are severely overestimated. This result
comes about because the Lorentzian factor decreases too slowly with 
depth at constant frequency. This is particularly so for modes of 
lowest frequency, for which the Lorentzian is widest.
Since it is not obvious where in the frequency
spectrum the transition from a Lorentzian (largest most-energetic eddies) to 
a Gaussian (small-scale turbulence) frequency spectrum takes place, 
results are shown only for the purely Gaussian frequency factor.

\vspace{-3mm}
\section{Solar-like amplitudes in other stars}
\vspace{-3mm}
Fairly accurate measurements of solar-like oscillation amplitudes
in other stars are available today from ground based observations 
(e.g. Kjeldsen et al. 2005; Bedding, these proceedings). 
Estimates of solar-like oscillation amplitudes in other stars were 
carried out by Christensen-Dalsgaard \& Frandsen (1983), 
Kjeldsen \& Bedding (1995), Houdek et~al. (1999), 
Houdek \& Gough (2002) and Samadi~et~al.~(2005).

\begin{table*}
\begin{center}
\caption{Comparison of scaling laws and theoretical velocity
         estimates with the observations of solar-like oscillations. The 
	 quantity $\hat V_{\rm obs}$ is the typical observed maximum apparent
	 velocity amplitude (defined to be $\sqrt2$ times the rms value),
         and $\hat V$ is the value obtained by the 
	 excitation theory, scaled to render $\hat V=\hat {\rm V}_\odot$ 
         for the Sun.
	 The solar value was taken to be $\hat{\rm V}_\odot$=\,0.23\,m\,s$^{-1}$
         (from Houdek \& Gough 2002).
        }
\bigskip
\renewcommand{\arraystretch}{1.2}
\begin{tabular}[t]{lccccccc}
\hfil Star&$M/{\rm M}_\odot$&$L/{\rm L}_\odot$&$T_{\rm e}/{\rm T}_{{\rm e}\odot}$
&$L{\rm M}_\odot/{\rm L}_\odot M$&${\rm g}_\odot/g$~&$\hat V/\hat{\rm V}_\odot$
&$\hat V_{\rm obs}/\hat{\rm V}_\odot$\\
\noalign{\medskip}
\hline
\noalign{\medskip}
$\alpha$\,Cen\,A&1.16&1.58~~&1.004&1.35~~~~~&1.34~~&1.39~~&1.5$^a$\\
$\beta$\,Hyi    &1.11&3.50~~&1.004&3.15~~~~~&3.11~~&3.25~~&2.2$^b$\\
Procyon         &1.46&6.62~~&1.107&4.53~~~~~&3.02~~&6.47~~&2.6$^c$\\
$\xi\,$Hya      &3.31&60.00~~&0.857&18.13~~~~~~~&33.64~~~&9.04~~&8.7$^d$\\
\end{tabular}
\end{center}
\hbox{\hspace{62pt} $^a$from Bouchy \& Carrier (2001),\quad
       $^b$from Bedding et al. (2001),\hfil}
\hbox{\hspace{62pt} $^c$from Martic et al. (1999),\quad\qquad\,
       $^d$Frandsen et al. (2002).}
\end{table*}
\begin{figure}[t]
\centering
\includegraphics[width=1.00\linewidth]{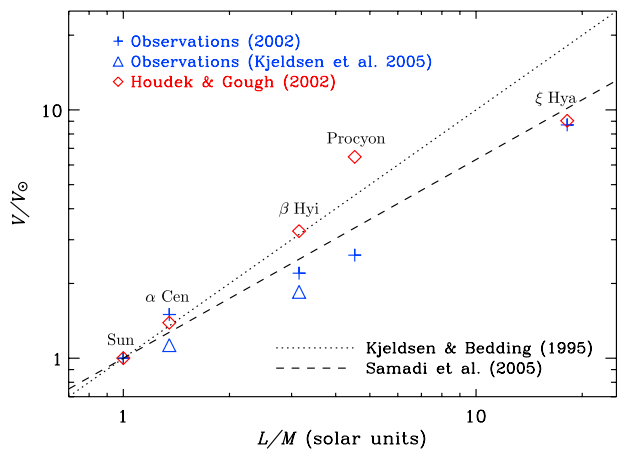}
\caption{
Predicted apparent velocity amplitudes (defined to be $\sqrt2$ times the rms 
value) as function of light-to-mass ratio for stochastically excited 
oscillations
in other stars. Observations from several authors are plotted by the
plus and triangle symbols. The theoretical estimates by Houdek \& Gough (2002)
are plotted as diamond symbols. The scaling law by Kjeldsen \& Bedding (1995) 
is illustrated by the dotted line and results reported by 
Samadi et~al. (2005) are indicated by the dashed line.
} 
\label{fig:veloamp}
\end{figure} 
\begin{figure}[t]
\centering
\includegraphics[width=1.00\linewidth]
                {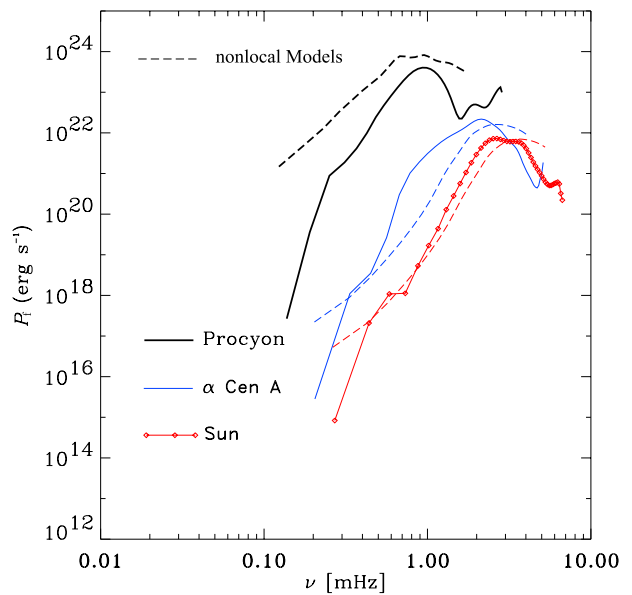}
\caption{
Acoustic excitation rates $P_{\rm f}$ of radial modes for the Sun
and for the two solar-type stars $\alpha$~Cen~A and Procyon.
The solid curves are the results from numerical simulations by 
Stein et~al. (2004) and the dashed curves are the model results
of Sections~3 and 6. The model results have been scaled with a factor 
$\Lambda$ such as to obtain the same maximum solar value of $P_{\rm f}$
than that from the numerical simulations (adapted from Stein et~al. 2004).
}
\label{fig:hydrosim}
\end{figure} 
As in the solar case, damping rates for other stars are obtained from
solving the eigenvalue problem~(\ref{eq:pulsation}) and from calculating
the excitation rates from expression~(\ref{eq:excitation-rate}).
With these estimates for $\eta$ and $P_{\rm f}$ Houdek \& Gough (2002)
predicted velocity amplitudes for several stars, using 
equation~(\ref{eq:V-H}). Results for stochastically
excited oscillation amplitudes and linear damping rates in the solar-like
star $\alpha$\,Cen\,A and in the sub-giant $\xi$\,Hydrae are illustrated
in Figs\,\ref{fig:aCenA} and \ref{fig:xiHya}.\;Bedding\;et\,al.\;(2004) reported
mode lifetimes for $\alpha$ Cen A between 1--2 days (see also
Fletcher, these proceedings) which are in 
reasonable agreement with the theoretical estimates of about $1.7$ days
for the most prominent modes 
(the mode lifetime $\tau\!=\!\eta^{-1}$; see lower panel of 
Fig.~\ref{fig:aCenA}).
For $\xi$~Hydrae, however, the theoretical mode lifetimes of the most 
prominent modes are $\tau\simeq 17$ days which are in stark contrast 
to the measured values of about 2--3 days by Stello et~al. (2004, 2006), 
yet the estimated velocity amplitudes for $\xi\,$ Hydrae are in almost 
perfect agreement with the observations by Frandsen et~al. (2002).
The estimated velocity amplitudes, together with theoretical results for 
other solar-like oscillators, are compared with measurements from
various observing campaigns and with the suggested scaling laws 
by Kjeldsen \& Bedding (1995, 2001) in Table~2.
For the cooler stars the theoretical results are in reasonable agreement
with the observations. For the rather hotter star Procyon, however, the
theoretical velocity amplitudes are severely overestimated (see Table 2).\\
This comparison between predicted and observed velocity amplitudes is
illustrated in Fig.~\ref{fig:veloamp}. The dotted line is the scaling
law by Kjeldsen \& Bedding (1995), and the dashed line is the scaling relation
reported by Samadi et~al. (2005) using the convective velocity profiles 
from numerical simulations (Stein \& Nordlund 2001), a
Lorentzian frequency factor in equation~(\ref{eq:function-S}), and the
theoretical damping rates from Houdek et~al. (1999). For hotter 
stars they find better agreement with observations.
\vspace{-5mm}
\section{Comparison with numerical simulations}
\vspace{-5mm}
Recently Stein et~al. (2004) reported stochastic excitation rates for radial
p modes in various stars obtained from hydrodynamical simulations of their 
surface convection zones. By comparing the excitation rates between these 
simulation results and the model computations discussed in the 
previous sections, we can put some constraints on the modelled oscillation 
amplitudes. In Fig.~\ref{fig:hydrosim} the excitation rates for the Sun
and for the two solar-like pulsators $\alpha$~Cen~A and Procyon are plotted
versus frequency. The solid curves are the results from the numerical
simulations and the dotted curves are the excitation rates obtained from
equation~(\ref{eq:excitation-rate}). The theoretical values were multiplied
with a scaling factor $\Lambda$ such as to obtain for the solar model the
same maximum value of $P_{\rm f}$ than that from the numerical simulations.
This allows us to compare the results for Procyon and $\alpha$~Cen~A between
the simulation ($P_{\rm sim}$) and model ($P_{\rm model}$) computations 
(from here on we omit the subscript f).
\begin{figure*}[t]
\centering
\includegraphics[width=1.0\linewidth]{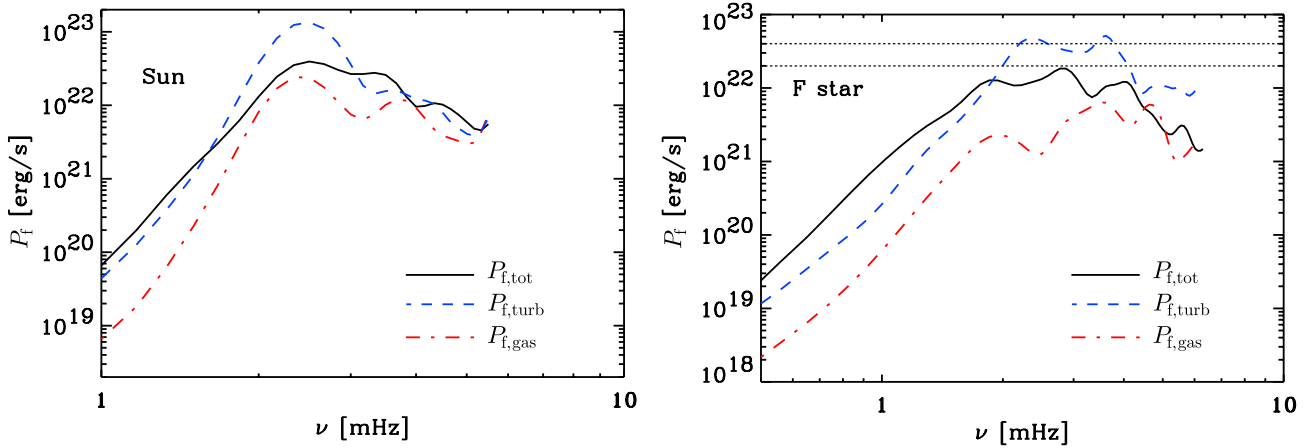}
\caption{
Total excitation rate $P_{\rm f, tot}$ and individual contributions from the
turbulent $P_{\rm f, turb}$ and gas $P_{\rm f, gas}$ pressure fluctuations 
obtained from numerical simulations for a solar model (left panel) and for a
model of an F star (right panel). For the solar model the two 
individual excitation sources are roughly similar in magnitude but for the
hotter F star the driving from the Reynolds stress (turbulent pressure) 
fluctuations dominates.
Partial cancellation between the Reynolds stress and gas pressure (buoyancy)
fluctuations leads to a total excitation rate $P_{\rm f, tot}$ that is 
smaller (by a factor of about two in the F star, indicated by the two 
horizontal, dotted lines) than the contribution from the Reynolds stresses
$P_{\rm f, turb}$ (adopted from Stein et~al. 2004).} 
\label{fig:exci-source}
\end{figure*} 
From this comparison (see Fig.~\ref{fig:hydrosim}) we obtain for Procyon
\begin{equation}
P_{\rm model}\simeq2\times P_{\rm sim}\,,
\label{eq:procyon_P}
\end{equation}
and from Table~2 we obtain for the ratio between the modelled ($V_{\rm model}$)
and observed ($V_{\rm obs}$) velocity amplitude for Procyon
\begin{equation}
V_{\rm model}\simeq2.5\times V_{\rm obs}\,.
\label{eq:procyon_V}
\end{equation}
With the assumption that the numerical simulations provide the ``correct''
maximum value for the excitation rate $P=P_{\rm sim}$, we can estimate
from the expressions $V=\sqrt{P/2\eta I}$ (see equation~(\ref{eq:V-H})), 
(\ref{eq:procyon_P}) and (\ref{eq:procyon_V}) the amount by which the 
modelled values of $P_{\rm model}$ and of $(2\eta I)^{-1}$ are 
at variance with the observations: 
\begin{equation}
V_{\rm model}\simeq\sqrt{2}\times1.77\times V_{\rm obs}\,,
\end{equation}
suggesting that the modelled excitation rate $P_{\rm model}$ is too large
by a factor of about two and the modelled value of $\sqrt{2\eta I}$ too small 
by a factor of about 1.77. In Section~4 we concluded that the modelled 
eigenfunctions, and therefore mode inertia $I$, reproduce the observations
reasonably well. Consequently we can argue that it is predominantly the
theoretical damping rate $\eta$ in $\sqrt{2\eta I}$ which contributes to the 
factor of 1.77, i.e. $\eta$ is too small (or the modelled mode lifetime 
too large).\\
A similar analysis can be done with $\alpha$~Cen~A, for which we obtain
from Fig.\ref{fig:hydrosim} and Table~2:
\begin{equation}
V_{\rm model}\simeq\sqrt{0.8}\times1.04\times V_{\rm obs}\,,
\end{equation}
suggesting that for this star the modelled damping rates $\eta$ are 
in good agreement with the observations (see also lower panel of 
Fig.\ref{fig:aCenA}).

It is interesting to note that the numerical simulations by 
Stein et~al. (2004) show partial cancellation between the 
two excitation sources ${\boldsymbol{\cal F}}(\boldsymbol u)$ and 
${\boldsymbol{\cal G}}(s^\prime)$ 
(see equation~(\ref{eq:excitation})) arising from the fluctuating
turbulent pressure (Reynolds stresses) and gas pressure (buoyancy force);
cancellation between acoustic multipole sources
was discussed before by Goldreich \& Kumar (1990) and Osaki (1990).
This effect is illustrated in Fig.~\ref{fig:exci-source} which shows the 
total excitation rate $P_{\rm f,tot}$ and the individual contributions
$P_{\rm f, turb}$ and $P_{\rm f, gas}$ from the turbulent and gas pressure
fluctuations obtained from numerical simulations of the surface convection
in the Sun and in a hotter star of spectral type F, the latter having model 
properties that are similar to that of Procyon. 
It is particularly striking that for the F star the
effect of this partial cancellation leads, for the most 
prominent modes, to a total excitation rate $P_{\rm f,tot}$ which is,
on average, by a factor of about two smaller 
than the excitation rate from only the turbulent pressure fluctuations 
$P_{\rm f, turb}$ (illustrated by the two dotted horizontal lines).
One is therefore tempted 
to argue that the overestimated values of the modelled excitation rate 
$P_{\rm f}$ in Procyon (which is an F5 star) could be partially attributed 
to having neglected the gas pressure fluctuations in 
equation~(\ref{eq:excitation}) and more importantly its cancellation 
effect with the turbulent pressure fluctuations. Including a formulation 
of this cancellation effect into the excitation model of Section 6.1 
could bring the estimated velocity amplitudes, particularly for hotter stars, 
in better agreement with the observations.
\vspace{-4.5mm}
\section*{Acknowledgements}
\vspace{-4.5mm}
I am grateful to Bill Chaplin for providing the top panels of 
Figs~\ref{fig:damping} and \ref{fig:heights}, to Hans-G\"unter Ludwig
and Regner Trampedach for providing their solar simulation results for
Fig.~\ref{fig:reynolds-stress} and to Douglas Gough for many helpful 
discussions. Support from the Particle Physics and Astronomy Research 
Council is gratefully acknowledged.


%
\end{document}